\documentclass[12pt,a4]{article}

\usepackage{amsmath} 
\usepackage{epsfig}
\usepackage{floatflt} 
\usepackage{subfigure}
\usepackage{theorem} 
\usepackage{url}
\usepackage{algorithmic}

\newcommand{\bra}[1]{\langle#1|}

\newcommand{\ket}[1]{|#1\rangle} 
\newcommand{\braket}[2]{\langle#1|#2\rangle}
\newcommand{\ketbra}[2]{|#1\rangle\langle#2|} 
\newcommand{\interm}[0]{|X|^2}
\newcommand{\poi}[2]{e^{-{#2}}\frac{{#2}^{#1}}{{#1}!}}  
\newcommand{\poit}[2]{p(#1;#2)}
\newcommand{\poitw}[3]{p_{#3}(#1;#2)}
\newcommand{\amp}[2]{a(#1;#2)}

\newtheorem{theorem}{Theorem} 

\newtheorem{proposition}{Proposition} 
\newtheorem{definition}{Definition}
\newtheorem{corollary}{Corollary}
\newcommand{\qed}{\hfill \mbox{\raggedright \rule{.07in}{.1in}}}
\newenvironment{proof}{\vspace{1ex}\noindent{\bf Proof}\hspace{0.5em}}
{\hfill\qed\vspace{1ex}\linebreak}

\sloppypar
\begin{document}

\title{Probability Ranking in Vector Spaces}
\author{Massimo Melucci\\\normalsize{University of Padua}}
\date{}
\maketitle
\begin{abstract}
  The Probability Ranking Principle states that the document set with the highest values
  of probability of relevance optimizes information retrieval effectiveness given the
  probabilities are estimated as accurately as possible. The key point of the principle is
  the separation of the document set into two subsets with a given level of fallout and
  with the highest recall. The paper introduces the separation between two vector
  subspaces and shows that the separation yields a more effective performance than the
  optimal separation into subsets with the same available evidence, the performance being
  measured with recall and fallout. The result is proved mathematically and exemplified
  experimentally.
\end{abstract}

\section{Introduction}
\label{sec:introduction}

Information Retrieval (IR) systems decide about the relevance. The decision about
relevance is subject to uncertainty.  A probability theory provides a measure of
uncertainty.  To this end, a probability theory defines the event space and the
probability distribution. The research in probabilistic IR is based on the classical
theory, which describes events and probability distributions using, respectively, sets and
set measures, according to Kolmogorov's axioms~\cite{Kolmogorov56}.

Ranking is perhaps the most crucial IR task and the Probability Ranking Principle (PRP)
reported in~\cite{Robertson77} is by far the most important theoretical result to date.
Although IR systems reach good results thanks to (classical) probability theory and
parameter tuning, ranking is far from being perfect because useless units are often ranked
at the top of, or useful units are missed from the retrieved document list.

The paper investigates whether an alternative probability theory may achieve further
improvement.  We propose Vector Probability to describe events and probability
distributions using vectors, matrices and operators on them. The adoption of Vector
Probability means a radical change: Vector Probability is based on vector subspaces
whereas classical probability is based on sets such that the regions of acceptance and
rejection of a hypothesis system are sets. We express Vector Probability by means of the
mathematical apparutus used in Quantum Theory. However, the use of the mathematical
apparatus of Quantum Theory does not end in an investigation or assertion of quantum
phenomena in IR. Rather, we argue the superiority of vector probability for ranking
documents over the classical probability theory.  Every reflection on Quantum Theory and IR
is out of the scope of the paper.

The paper shows that ranking in accordance with Vector Probability is more effective than
ranking by classical probability given that the same evidence is available for probability
estimation. The effectiveness is measured in terms of probability of correct decision or,
equivalently, of probability of error. We propose a decision rule based on vector
subspaces such that, in the long run, the IR system will deem a document relevant
correctly at a higher recall than the recall measured in the event of ranking as a result
of classical probability when the fallout is not more than a given threshold. So, the
decision rule minimizes the probability of error.

We organize the paper as follows. Section~\ref{sec:definitions} provides the definitions
used in the paper: this section can be skipped if the reader has knowledge about Quantum
Theory and Probability; further definitions can be found in~\cite{vanRijsbergen04}.
Section~\ref{sec:prob-relev} introduces the aspects of the probability of relevance
related to the subsequent sections.  Section~\ref{sec:vect-relev-prob} explains Vector
Probability.  Section~\ref{sec:relev-prob-repr} describes an instance of the vector
probability of relevance when the Poisson distribution is used for an observable of a
document.  Section~\ref{sec:optim-observ} introduces the optimal observable vectors.
Section~\ref{sec:optim-prob-rank} shows that ranking by means of the optimal observable
vectors is more effective than ranking by means of the best subsets of observed
values. Section~\ref{sec:experiments} describes an experimental study that confirms the
theory.  Section~\ref{sec:discussion} is about the measurement of observable vectors and
makes some remarks about the actual use.  Section~\ref{sec:related-work} refers to the
main related publications and Section~\ref{sec:conclusions} concludes the paper.

\begin{table*}[t]
  \centering
    \footnotesize
  \begin{tabular}[t]{|l|l|}
   \hline
    Definition			& IR Example or Corresponding Concept			\\
    \hline
    Observable			& Term frequency, relevance, color, aboutness		\\
    Probability Distribution	& Distribution of probability of term frequency		\\
    State			& Relevance, aboutness, utility				\\
    Probability of Detection	& Recall						\\
    Probability of False Alarm	& Precision						\\
    Region of Acceptance	& Term frequencies that are higher than a threshold	\\
    Observable Vector		& Term frequency, relevance, color, aboutness		\\
    State Vector		& Distribution of Relevance, aboutness, utility	probability	\\
    \hline
  \end{tabular}
  \caption{Definitions, IR examples and concepts.}
  \label{tab:definitions}
\end{table*}

\section{Definitions}
\label{sec:definitions}

We report and compare some definitions to IR concepts in Table~\ref{tab:definitions}.
\begin{definition}[Observable]
  An observable is a property that can be measured from an entity.
\end{definition}
\begin{definition}[Probability Distribution]
  A probability distribution is a function that maps observable values to the real range
  $\left[0, 1\right]$. As usual, the probabilities sums to $1$.
\end{definition}
\begin{definition}[Classical Probability Distribution]
  A classical probability distribution admits  only  \emph{sets} of observable values.
\end{definition}
The subsets of observable values can be defined by means of the
set operations (i.e., intersection, union, complement).
\begin{definition}[State]
  A state, or hypothesis, is a condition of the measured entity and molds the probability
  distribution of the measurement. 
\end{definition}
In classical probability, ``hypothesis'' is more used than ``state''.  We use ``state''
because it is used in the formalism of Quantum Theory. We correspond the null state to
non-relevance and the alternative state to relevance. An IR system decides between the
relevance state and the non-relevance provided an observable value.
\begin{definition}[Probability of Detection]
  It is the probability that the system decides for relevance when relevance is true; it
  is also called power.
\end{definition}
\begin{definition}[Probability of False Alarm]
  It is the probability that the system decides for relevance when relevance is false; it
  is also called size or level.
\end{definition}
As the probability of detection and the probability of false alarm cannot be
simultaneously optimized, the decision rules maximize the probability of detection when
the probability of false alarm is fixed.
\begin{definition}[Region of Acceptance]
  A region of acceptance consists of the observable values that induce the
  system to decide for relevance. The most powerful region of acceptance yields the
  maximum power for a fixed size. 
\end{definition}
Note that ``acceptance'' does often refer to the null state in Statistics. 

The Neyman-Pearson lemma states that the maximum likelihood (ML) ratio test defines the
most powerful region of acceptance~\cite{Neyman&33}.
\begin{definition}[Probability of Correct Decision]
  \begin{equation}
    \label{eq:18}
    P_c = \xi (1-P_0) + (1-\xi)P_d
  \end{equation}
  provided that $\xi$ the prior probability of the null state, $P_0$ is the probability of false
  alarm and $P_d$ is the probability of detection.
\end{definition}
\begin{definition}[Probability of Error]
  \begin{equation}
    \label{eq:20}
    P_e = \xi P_0  + (1-\xi) (1-P_d)
  \end{equation}
\end{definition}
From~\ref{eq:18} and~\ref{eq:20}, $P_e+P_c=1$.
\begin{definition}[Vector Space]
  A vector space over a field $\cal F$ is a set of vectors subject to linearity, namely, a
  set such that, for every vector $\ket{u}$, there are three scalars $a,b,c \in \cal F$
  and three vectors $\ket{v}, \ket{x}, \ket{y}$ of the same space such that $\ket{u} =
  a\ket{v}$ and $\ket{u} = b\ket{x}+c\ket{y}$. If $\ket{u}$ is a vector, $\bra{u}$ is its
  transpose, $\braket{v}{u}$ is the inner product with $\ket{v}$ and $\ketbra{v}{u}$ is
  the outer product with $\ket{v}$. If $\left|\braket{x}{x}\right|^2 = 1$, the vector is
  normal. If $\braket{x}{y} = 0$, the vectors are mutully orthogonal.
\end{definition}
We adopt the Dirac notation to write vectors so that the reader may refer to the
literature on Quantum Theory; a brief illustration of the Dirac notation is
in~\cite{vanRijsbergen04}.
\begin{definition}[Observable and Observable Vector]
  An observable is a collection of values and of vectors. The observable vectors are
  mutually \emph{orthonormal} and 1:1 correspondence with the values.
\end{definition}
An observable corresponds to a random variable in Statistics whereas the observable
vectors correspond to the indicator functions.
\begin{definition}[State Vector]
  A state vector defines a vector probability distribution. The possible states (or
  hypotheses) correspond to state vectors.
\end{definition}
A state vector plays the role of parameters in Statistics.
\begin{definition}[Vector Probability]
  The vector probability that ${x}$ is observed given the state ${m}$ is
  $\left|\braket{x}{m}\right|^2$.
\end{definition}
Vector probability is axiomatically defined in~\cite{vonNeumann55} and is applied to IR
in~\cite{vanRijsbergen04}; the generalization to state matrices or density matrices is not
necessary in this paper.

\section{Probability of Relevance}
\label{sec:prob-relev}

Suppose that a document is represented through the random variable $X$ such that $X=x$ means
that a term has frequency $x$.  The decision is between relevance and non-relevance. Thus,
the probability of detection is the probability that the observed frequency belongs to the
region of acceptance when a document is relevant and the probability of false alarm is the
probability that the observed frequency belongs to the region of acceptance when the document
is not relevant.

The very general form of Probability Ranking Principle (PRP) and then the BM25 reported
in~\cite[page 340]{Robertson&09} are based on the ML ratio test.  The PRP states that, if
a cut-off is defined for the fallout (i.e., the probability of false alarm), we would
clearly optimize (i.e., maximize recall, namely, probability of detection, or
equivalently, precision) if we included in the retrieved set those documents with the
highest probability of relevance~\cite[page 297]{Robertson77} which is the probability
that $X=x$ when the document is relevant.

Therefore, the PRP and the Neyman-Pearson lemma state that, given a region of acceptance
and then a probability of false alarm (i.e., fallout), the document ranking as a result of probability
of relevance is optimal because the recall is maximum.

\section{Vector Probability of Relevance}
\label{sec:vect-relev-prob}

Suppose that $X$ is an observable (e.g., term frequency) and $x$ a value of the set
$\left\{{0}, {1}, \ldots, {N}\right\}$. The orthonormal observable vectors that correspond
to the values are $\ket{0}, \ket{1}, \ldots, \ket{N}$; the actual implementation of these
vectors is not essential.  A observable vector $\ket{x}$ correspond to $x$ and
\begin{equation}
  \label{eq:4} X = \sum_{x=0}^N x \ketbra{x}{x}
\end{equation}  

Suppose that $\poit{x}{m}$ is the probability that $X=x$ given a parameter $m$. In the
event of binary relevance, $m$ is either $m_0$ (non-relevance) or $m_1$ (relevance). Note
that $m$ may refer to more than one parameter. However, we assume that $m$ is scalar for
the sake of clarity.

Two relevance state vectors represent binary relevance: a relevance state vector
$\ket{m_0}$ represents non-relevance state and an orthogonal relevance state vector
$\ket{m_1}$ represents relevance state.  Relevance state vectors and the observable
vectors belong to an finite-dimensional vector space.\footnote{In Quantum Theory, the
  vector spaces are complex Hilbert spaces. For the sake of simplicity, we do not consider
  the field.} Thus, either state vectors can be defined in terms of a given orthonormal
basis of that space and, in particular, the observable vectors are a basis.  The following
expressions
\begin{eqnarray}
  \ket{m_0} = \sum_{x=0}^{N} \amp{x}{m_0} \ket{x} \nonumber\\ 
  \ket{m_1} = \sum_{x=0}^{N} \amp{x}{m_1} \ket{x}\nonumber\\
  \amp{x}{m} = \pm \sqrt{\poit{x}{m}}
  \label{eq:5} 
\end{eqnarray}
establish the relationship between parametric distributions and vector spaces, namely,
between the parameters $m_0, m_1$, the relevance state vectors $\ket{m_0}, \ket{m_1}$ and
the observable $X$.  The sign of $\amp{x}{m}$ is chosen so that the orthogonality between
the state vectors is retained.  Equations~\ref{eq:5} are instances of superposition. In
Physics, superposition models observables that are known only if they are measured.  In
IR, the event that an observable exists only if observed is a much debated hypothesis.
Moreover, the orthogonality of the relevance state vectors and the following expression
\begin{eqnarray}
  \label{eq:9}
  \left|\braket{y}{m}\right|^2&=& \left|\sum_{x=0}^N \amp{x}{m}  \braket{y}{x}\right|^2\\
  {}			      &=& \left|\amp{y}{m}\right|^2 \qquad \mbox{due to orthogonality}\\
  {}			      &=& \poit{y}{m}
\end{eqnarray}
establish the relationship between probability distribution and vector-based
representation of relevance.

\section{Poisson-Based Probability of Relevance}
\label{sec:relev-prob-repr}

The Poisson distribution is used because we want to make the illustration of the theory
accessible in the remainder of the section and consistent with the past literature,
e.g.,~\cite{Harter75a,Harter75b,Robertson&94,Robertson&81}.  Moreover, the Poisson
distribution is asymptotically derived from the Binomial distribution and approximates the
Normal distribution.

The observable $X$ is the frequency of a term in a document. Thus, $X = x$ means that the
term occurs $x$ times in a document.  The Poisson probability distribution gives the
probability that $X=x$, that is,
\begin{equation}
  \label{eq:3} \poit{x}{m}=\poi{x}{m}
\end{equation} 
provided that $m$ is the expected term frequency. $X$ is defined in the set of natural number.
However, we assume that $N$ is finite, large and equal to the maximum observable term
frequency in the collection; indeed, the estimated probability that a term frequency is
greater than $N$ is zero.


Two distinct parameters $m_0, m_1$ encode non-relevance and relevance, respectively, in
parametric Statistics,. Thus,
\begin{equation}
  \poit{x}{m_0}\label{eq:7}
\end{equation}
is the probability that the term occurs $x$ times in a \emph{non-relevant} document and
\begin{equation}
  \poit{x}{m_1}\label{eq:8}
\end{equation}
is the probability that the term occurs $x$ times in a \emph{relevant} document. 

\begin{table}[t]
  \centering
  \begin{tabular}[t]{|c|c|c|}
    \hline
    {}			&  		&              	\\[-6pt]
    {}			& $\poit{0}{m}$	& $\poit{1}{m}$	\\[2pt]
    \hline
    {}			&  		&              	\\[-6pt]
    $m_0$		& $\frac{1}{5}$	& $\frac{4}{5}$	\\[2pt]
    $m_1$		& $0$		& $1$		\\[2pt]
    \hline
  \end{tabular}
  \label{tab:example-0}
\end{table}
For example, consider the probability distributions in Table~\ref{tab:example-0}. We have
that
\begin{eqnarray*}
  && X = 0\ketbra{0}{0} + 1\ketbra{1}{1} = \ketbra{1}{1}\\
  && \ket{m_0} = \frac{1}{\sqrt{5}} \ket{0} + \frac{2}{\sqrt{5}} \ket{1} \qquad \ket{m_1} = \ket{1}\\
  && \poit{0}{m_0} = \left|\amp{0}{m_0}\right|^2 = \frac{1}{5} \qquad 
  \poit{0}{m_1} = \left|\amp{0}{m_1}\right|^2 = 0
\end{eqnarray*}

The Poisson-based probabilities of detection and false alarm are, respectively,
\begin{equation}
  \label{eq:10}
  P_d = \sum_{x=x_\alpha}^N \poit{x}{m_1} \qquad P_0 = \sum_{x=x_\alpha}^N \poit{x}{m_0}
\end{equation}
assuming that $N$ is so large that $\poit{x}{m} = 0, x > N$ and $\left\{x_\alpha, \ldots,
  N\right\}$ is the region of acceptance of the state of relevance at \emph{size}
$\alpha$.

When the states are equiprobable and $m_1 > m_0$, the Poisson-based probability of
error and the Poisson-based probability of correct decision are
\begin{eqnarray}
  \label{eq:12}
  P_e &=& \frac{1}{2}\left(\sum_{x=x_\alpha}^N \poit{x}{m_0} + \sum_{x=0}^{x_\alpha-1} \poit{x}{m_1}\right)\\
  {}  &=& \frac{1}{2}\left(1-\int_{m_0}^{m_1} \frac{t^{x_\alpha-1}e^{-t}}{\Gamma(x_\alpha)}dt\right)\\
  P_c &=& \frac{1}{2}\left(\sum_{x=x_\alpha}^N \poit{x}{m_1} + \sum_{x=0}^{x_\alpha-1} \poit{x}{m_0}\right)\\
  {}  &=& \frac{1}{2}\left(1+\int_{m_0}^{m_1} \frac{t^{x_\alpha-1}e^{-t}}{\Gamma(x_\alpha)}dt\right)
\end{eqnarray}
The greater the difference between $m_0$ and $m_1$, the greater $P_c$ and the smaller
$P_e$. If $m_1 < m_0$, the superscript and the subscript of the integral function
of~\eqref{eq:12} swap. If $m_1 = m_0$, error and correct decision are equiprobable, i.e.,
the decision is ruled through coin tossing.  In other words, the discrimination power increases
when $|m_1 - m_0|$ increases.  Note that the increase of $|m_1 - m_0|$ corresponds to
making the relevance state vectors orthogonal.

Probability of error and probability of correct decision provide an alternative form of
the decision procedure.  From Equations~\ref{eq:18} and~\ref{eq:20}, the maximum $P_d$ and
the minimum $P_0$ minimize $P_e$ and maximize $P_c$. Suppose we have three size values:
$\alpha_0 \leq \alpha_1 \leq \alpha_2$, thus yielding three power values $\beta_0 \leq
\beta_1 \leq \beta_2$.  The prior probabilility minimizes the probability of error as
follows:
\begin{equation}
  \label{eq:25}
  \min P_e =
  \left\{
    \begin{array}{ll}
      \xi\alpha_0 + (1-\xi)(1-\beta_0) & 0 \leq \xi < \xi_1\\
      \xi\alpha_1 + (1-\xi)(1-\beta_1) & \xi_1 \leq \xi < \xi_2\\
      \xi\alpha_2 + (1-\xi)(1-\beta_2) & \xi_2 \leq \xi < 1
    \end{array}
  \right.
\end{equation}
such that
\begin{eqnarray*}
  \xi_1 &= \frac{\beta_1 - \beta_0}{\alpha_1 - \alpha_0 + \beta_1 - \beta_0} &= \frac{4}{7}
  \\ 
  \xi_2 &= \frac{\beta_2 - \beta_1}{\alpha_2 - \alpha_1 + \beta_2 - \beta_1} &= \frac{8}{23}
\end{eqnarray*}
\begin{figure}[t]
  \centering
  \epsfig{file=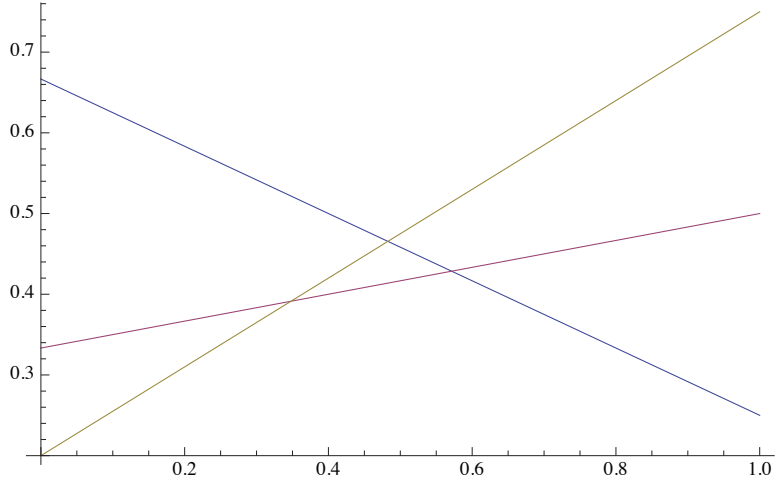}
  \caption{Polygonal curve that minimizes the probability of error. The polygonal curve is
  obtained from the lowest segments connected at the intersection points.}
  \label{fig:polygonal}
\end{figure}

Figure~\ref{fig:polygonal} shows an example of the polygonal curve with
$\alpha_0=\frac{1}{4}, \alpha_1 = \frac{1}{2}, \alpha_2 = \frac{3}{4}$ and $\beta_0 =
\frac{1}{3}, \beta_1 = \frac{2}{3}, \beta_2 = \frac{4}{5}$.  The abscissas of the
intersection points are $\xi_1, \xi_2$.

\section{Optimal Observable Vectors}
\label{sec:optim-observ}

Let us recapitulate some facts. Neyman-Pearson's lemma states that the set of term
frequencies can be partitioned into two disjoint regions: one region includes all the
frequencies such that relevance will be accepted; the other region denotes rejection. If a
term is observed from documents and only presence/absence is observed, the set of the
observable values is $\{0, 1\}$ and each region is one out of possible subsets, i.e.,
$\emptyset, \{0\}, \{1\}, \{0,1\}$. 

If term frequency is observed instead, the observable values are $\{0, 1, \ldots, N\}$ and
each region is one out of possible subsets of $\{0, 1, \ldots, N\}$. Note that the ML
ratio test yields two subsets, i.e., $\{0, \ldots, x_\alpha-1\}$ and $\{x_\alpha, \ldots,
N\}$. An alternative region can be defined with only set operations (intersection,
complement, union). However, set operations cannot define more powerful regions than that
dictated by dint of the Neyman-Pearson lemma.

The subspaces are equivalent to subsets and they then can be subject to set operations
\emph{if} they are mutually orthogonal~\cite{Griffiths02}. The subsets yielded by dint of
the ML ratio test become $\{\ketbra{0}{0}, \ldots, \ketbra{x_\alpha-1}{x_\alpha-1}\}$ and
$\{\ketbra{x_\alpha}{x_\alpha}, \ldots, \ketbra{N}{N}\}$ and can be subjected to set
operations because they are mutually orthogonal.

\begin{figure}[t]
  \centering
  \epsfig{width=0.75\columnwidth,file=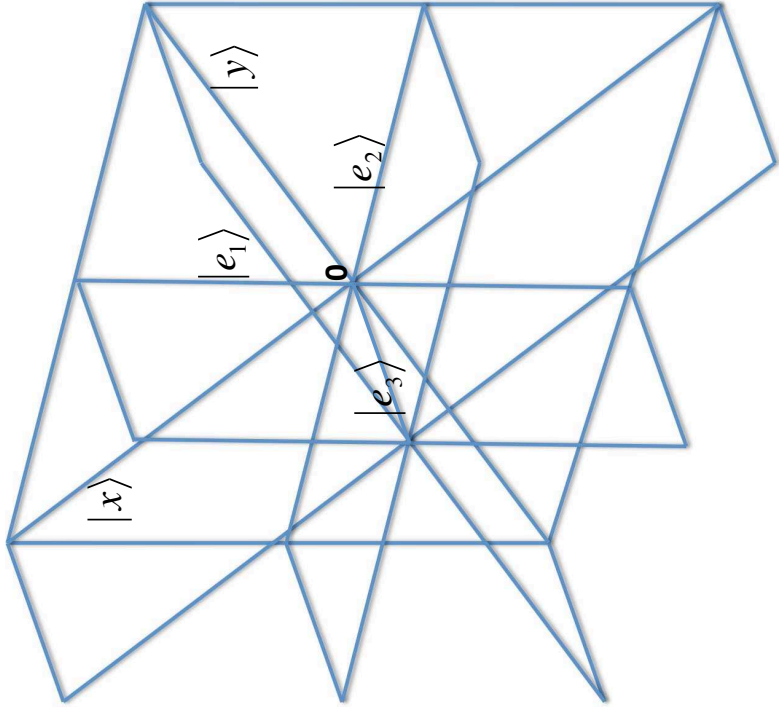}  
  \caption{The difference between subsets and subspaces.}
  \label{fig:subspaces}
\end{figure}
 
Suppose that the vector subspaces that correspond to the subsets yielded by dint of the ML
ratio test, are rotated so that the new subspaces are oblique to them.  The new oblique
subspaces cannot be reformulated in terms of the observable vectors through set operations
and thus they represent something different. 

Figure~\ref{fig:subspaces} shows three-dimensional vector space spanned by $\ket{e_1},
\ket{e_2}, \ket{e_3}$ to make the difference between subsets and subspaces. The ray (i.e.
one-dimensional subspace) $L_x$ is spanned by $\ket{x}$, the plane (i.e. two-dimensional
subspace) $L_{x,y}$ is spanned by $\ket{x}, \ket{y}$. Note that $L_{e_1,e_2} = L_{x,y} =
L_{e_1,y}$ and so on. According to~\cite[page 191]{Hughes89}, consider the subspace
$L_{e_2} \wedge (L_y \vee L_x)$ provided that $\wedge$ means ``intersection'' and $\vee$
means ``span'' (and not set union). Since $L_y \vee L_x = L_{x,y} = L_{e_1,e_2}$, $L_{e_2}
\wedge (L_y \vee L_x) = L_{e_2} \wedge L_{e_1,e_2} = L_{e_2}$. However, $(L_{e_2} \wedge
L_y) \vee (L_{e_2} \wedge L_x) = 0$ because $L_{e_2} \wedge L_y = 0$ and $L_{e_2} \wedge
L_x = 0$, therefore \begin{equation*}
  L_{e_2} \wedge (L_y \vee L_x) \neq (L_{e_2} \wedge L_y) \vee (L_{e_2} \wedge L_x)
\end{equation*}
thus meaning that the distributive law does not hold, hence, set operations cannot be
applied to subspaces.

The key point is that, if the subspaces are rotated in an optimal way, we can achieve the
most powerful regions; these regions cannot be ascribed to the subsets yielded by dint of
the ML ratio test. The following Helstrom's theorem is the rule to compute the most
powerful regions of a vector space provided two state vectors.

\begin{theorem}
  \label{the:helstrom}
  Let $\ket{m_1}, \ket{m_0}$ be the state vectors.  The region of acceptance at the highest
  probability of detection at every probability of false alarm is given by the
  eigenvectors of
  \begin{equation}
    \ketbra{m_1}{m_1} - \ketbra{m_0}{m_0}\label{eq:26}
  \end{equation}
  whose eigenvalues are positive.
\end{theorem}
\begin{proof}
  See~\cite{Helstrom76}. (The $\ketbra{m_i}{m_i}$ are defined in
  Section~\ref{sec:definitions}.)
\end{proof}
\begin{definition}
  An optimal observable vector is a vector that divides the region of acceptance from the
  region of rejection as stated by Theorem~\ref{the:helstrom}.
\end{definition}
The optimal observable vectors always exist due to the Spectral Decomposition theorem.~\cite{Halmos87}

The angle between the relevance state vectors $\ket{m_1}, \ket{m_0}$ determines the
geometry of the decision between the two states.  Suppose $\ket{\mu_1}, \ket{\mu_0}$ are
two observable vectors.  They are mutually orthogonal because are eigenvectors
of~\eqref{eq:26} and can be defined in the space spanned by the relevance state vectors.
The probability of correct decision and the probability of error are given by the angle
between the two relevance state vectors and by the angles between the observable vectors
and the relevance state vectors; how the geometry defines the optimal ranking is described
in the next section.

\begin{figure}[t] 
  \centering 
  \setlength{\unitlength}{1mm}  
  \subfigure[Non-optimal observable vectors are mutually orthogonal and
  \emph{asymmetrically}  placed around $\ket{m_0}, \ket{m_1}$]{
    \begin{minipage}[t]{1.0\columnwidth}
      \begin{picture}(50,60)(-25,-40)
        \put(0,0){\vector(3,+1){41}}\put(41,13){$\ket{m_1}$}
        \put(0,0){\vector(3,-1){41}}\put(41,-14){$\ket{m_0}$}
        \thicklines
        \put(0,0){\vector(2,+1){39}}\put(40,18){$\ket{e_1}$}
        \put(0,0){\vector(1,-2){19}}\put(20,-40){$\ket{e_0}$}
        \thinlines
        \put(0,0){\circle{100}}
        \put(8,-0.5){$\gamma$}
        \put(20,8){$\eta_1$}
        \put(20,-20){$\eta_0$}
      \end{picture}
   \end{minipage}
    \label{fig:geometry-a}} 
  \subfigure[Optimal observable vectors are mutually orthogonal and
  \emph{symmetrically}  placed around $\ket{m_0}, \ket{m_1}$]{
    \begin{minipage}[t]{1.0\columnwidth}
      \begin{picture}(50,65)(-25,-30)
        \put(0,0){\vector(3,+1){41}}\put(41,13){$\ket{m_1}$}
        \put(0,0){\vector(3,-1){41}}\put(41,-14){$\ket{m_0}$}
        \thicklines
        \put(0,0){\vector(1,+1){30}}\put(31,30){$\ket{\mu_1}$}
        \put(0,0){\vector(1,-1){30}}\put(31,-30){$\ket{\mu_0}$}
        \thinlines
        \put(0,0){\circle{100}}
        \put(8,-0.5){$\gamma$}
        \put(7,4){$\theta$}
        \put(7,-6){$\theta$}
      \end{picture}
    \end{minipage}
    \label{fig:geometry-b}}
  \caption{Geometry of decision and observable vectors}
  \label{fig:geometry}
\end{figure}
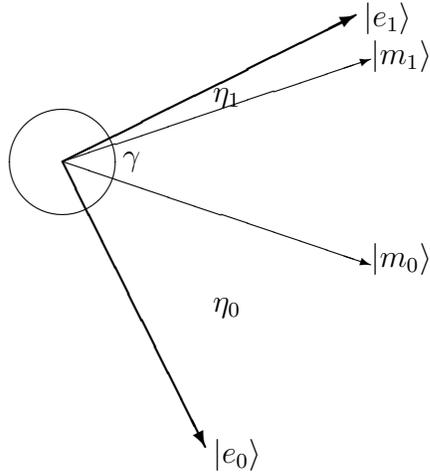
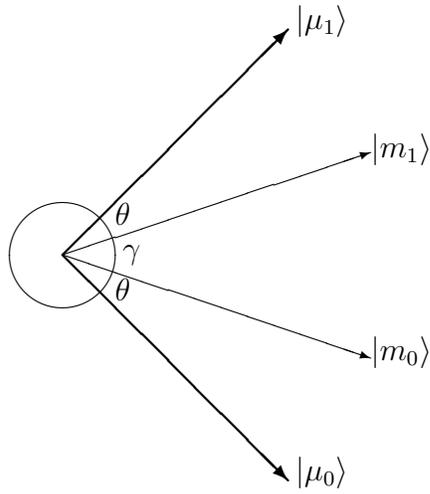

\section{Optimal Probability Ranking}
\label{sec:optim-prob-rank}

Figure~\ref{fig:geometry-a} illustrates the geometry of decision and the observable
vectors. (The figure is in the two-dimensional space for the sake of clarity, but the
reader should generalize to higher dimensionality than two.) The angles $\eta_0, \eta_1$
between the observable vectors and the relevance state vectors $\ket{m_0}, \ket{m_1}$ are
related with the angle $\gamma$ between $\ket{m_0}, \ket{m_1}$ because the observable
vectors are always mutually orthogonal and then the angle is $\frac{\pi}{2} = \eta_0 +
\gamma + \eta_1$.

The optimal observable vectors are achieved when the angles between an observable
vector and a relevance state vector are equal to
\begin{equation}
  \label{eq:1} 
  \theta = \frac{\frac{\pi}{2}-\gamma}{2}
\end{equation} 
The rotation of the non-optimal observable vectors such that~\eqref{eq:1} holds, yields
the optimal observable vectors $\ket{\mu_1}, \ket{\mu_0}$ as Figure~\ref{fig:geometry-b}
illustrates: the optimal observable vectors are ``symmetrically'' located around the
relevance state vectors.

The replacement of the angle between an observable vector and a relevance state vector
with the angle of~\eqref{eq:1} yields the minimal probability of error and the maximal
probability of correct decision, that is,
\begin{equation}
  \label{eq:2} Q_e = \frac{1}{2}\left(1-\sqrt{1-4\xi(1-\xi)\interm}\right)
  \qquad Q_c = 1-Q_e
\end{equation} 
(see~\cite{Helstrom76}) given that
\begin{equation}
  \label{eq:24}
  \interm = \left|\braket{m_0}{m_1}\right|^2 = \left|\sum_{x=0}^N \sqrt{\poi{x}{m_0}} \sqrt{\poi{x}{m_1}}\right|^2
\end{equation}
is the squared cosine of the angle between the relevance state vectors if the Poisson distribution is used.
\begin{figure}[t]
  \centering
    \epsfig{file=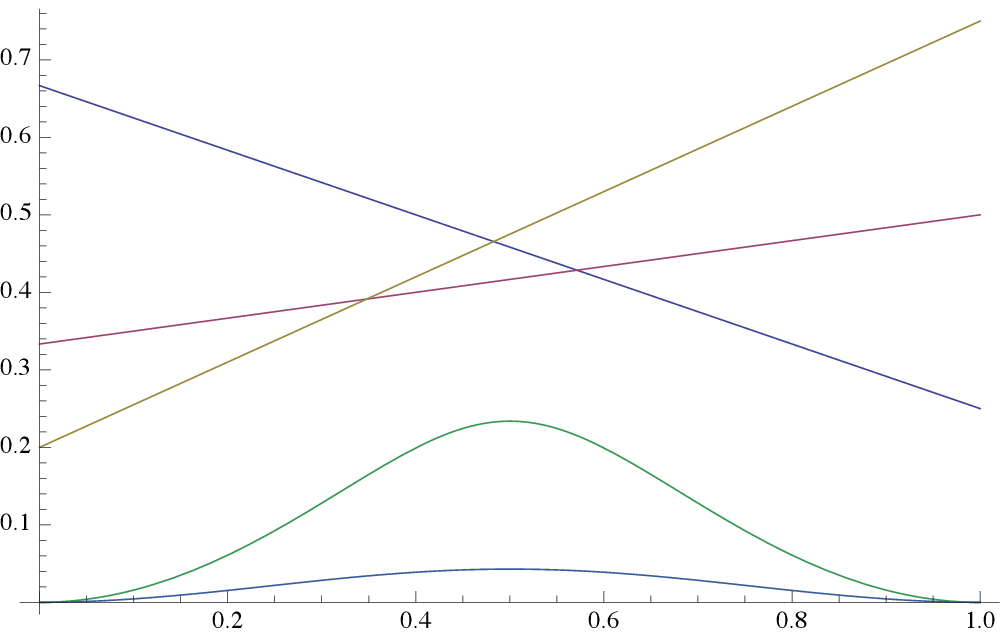}
    \caption{Probabilities of error: $P_e$ vs. $Q_e$.  Figure~\ref{fig:polygonal}
      describes the polygonal curve.  The shortest bell-shaped curve corresponds to $\xi =
      0.50$, whereas the other bell-shaped curve corresponds to $\xi = 0.90$.}
  \label{fig:probability-of-error}
\end{figure}
Figure~\ref{fig:probability-of-error} superposes the polygonal curve plotted for $P_e$ and
the bell-shaped curves plotted for $Q_e$ with $\interm=0.90$ and $\interm=0.50$.
$\interm$ measures the degree to which the distributions of probability of relevance and
non-relevance are distinguishable.  The less they are distiguishable, the higher $Q_e$.
Indeed, the probability of error increases when the distribution of probability of
relevance is very similar to the distributions of probability of non-relevance.

We prove the following
\begin{proposition}
  \label{sec:optim-observ-1}
  For all $m_0, m_1, x_\alpha$,
  \begin{equation}
    \label{eq:13}
    Q_c \geq P_c \qquad Q_e \leq P_e
  \end{equation}
\end{proposition}
\begin{proof}
  Let $x_\alpha \geq 0$ and let $m_1 \geq m_0$ -- the complement is proved in similar way.
 \begin{equation}
    \label{eq:14}
    Q_c \geq P_c \ \mbox{if and only if} \ \sqrt{1-\interm} \geq \int_{m_0}^{m_1}  \frac{t^{x_\alpha-1}e^{-t}}{\Gamma(x_\alpha)}dt
  \end{equation}
  because~\eqref{eq:18} and~\eqref{eq:20} also hold for $Q_c, Q_e$.
  Moreover,~\eqref{eq:14} holds if 
  \begin{equation}
    \label{eq:15}
    1-\interm \geq \int_{m_0}^{m_1} \frac{t^{x_\alpha-1}e^{-t}}{\Gamma(x_\alpha)}dt
  \end{equation}
  because the sides of~\eqref{eq:14} lies between $0$ and $1$.  The calculation of the
  angle between the relevance state vectors yields
  \begin{equation}
    \label{eq:6}
    \interm = e^{-\left|m_1-m_0\right|}\ .
  \end{equation}
  The relationships between the Poisson distribution and the Gamma function allows us to
  state that
  \begin{equation}
    \label{eq:17}
    1-e^{-\left|m_1-m_0\right|} = e^{-m_1}\sum_{x=0}^{\infty} \frac{m_1^x}{x!} -
    e^{-m_1}\sum_{x=0}^{\infty} \frac{m_0^x}{x!} 
  \end{equation}
  We split the summations in~\eqref{eq:17}, thus achieving that~\eqref{eq:15} holds if and
  only if
  \begin{eqnarray}
    \nonumber 
    2 \sum_{x=0}^{x_\alpha-1} \poi{x}{m_1} + \\ 
    \nonumber 
    e^{-m_1}\sum_{x=x_\alpha}^{N} \left(\frac{m_1^x}{x!} - \frac{m_0^x}{x!}\right) + \\ 
    \left( e^{-m_0} - e^{-m_1}\right) \sum_{x=0}^{x_\alpha-1} \frac{m_0^x}{x!} \geq 0
    \label{eq:19}
  \end{eqnarray}
  Every operand of the sum~\eqref{eq:19} is not negative, thus proving the left side
  of~\eqref{eq:13}.   The right side is proved in symmetric way due to~\eqref{eq:20} when
  applied to $Q_e, Q_c$.
\end{proof}
Proposition~\ref{sec:optim-observ-1} tells us that the decision as to whether a document
is relevant is most effective when the test is function of the optimal observable vectors
even if the Poisson-based probability is estimated as accurately as possible.
\begin{corollary}
  For all $m_0, m_1, x_\alpha$ and if $Q_0 = P_0$,
  \begin{equation}
    \label{eq:21}
    Q_d \geq P_d 
  \end{equation}
  \label{sec:optim-observ-2}
\end{corollary}
\begin{proof}
  Let $Q_0 = P_0$, recall~\eqref{eq:18},~\eqref{eq:20} and the left side
  of~\eqref{eq:13}.  Thus,
  \begin{equation}
    \label{eq:16}
    0 \leq Q_c - P_c = (1-\xi)(Q_d-P_d)
  \end{equation}
  We have that $Q_c \geq P_c$ because $Q_d \geq P_d$ and $0 \leq \xi \leq 1$.
\end{proof}
Corollary~\ref{sec:optim-observ-2} tells us that, once the probability of false
alarm is fixed at an arbitrary size, the state that a document is relevant is
correctly accepted with vector probability that is higher than any Poisson-based
probability.

The key point is that the region of acceptance induced by the optimal observable vectors
$\mu_1, \mu_0$ is more powerful than the region of acceptance of the PRP, when the Poisson
distribution measures the probability of term frequency, all the other things being
equal. A distribution different from Poisson's or a different estimation of the
probability values might revert the outcome of Corollary~\ref{sec:optim-observ-2}. Does
the power of the region of acceptance induced through the optimal observable vectors and then
Corollary~\ref{sec:optim-observ-2} depend on the probability of term frequency?  In the
remainder of the section, we generalize the result for either distribution of probability
of term frequency.

The probability of term frequency is given by two items: (\emph{i}) the probability values
estimated for each $x$; (\emph{ii}) the distribution used to compute the probability of
term frequency.

As for (\emph{i}), note that the vector probability and the classical probability of
relevance are functions of the same probability distributions $p(x,m_0), p(x,m_1)$ for
every $m_0,m_1$.  Thus, the power of the region of acceptance induced through the optimal
observable vectors and then Corollary~\ref{sec:optim-observ-2} do not depend on the
probability values calculated for each $x,m$.

As for (\emph{ii}), what distinguishes the probability of detection (and false alarm)
computed with the optimal observable vectors from those computed with the region of
acceptance given as a result of the PRP (i.e., the ML ratio test) is the region of acceptance.  
We prove that the region of acceptance given through the optimal observable vectors is always
more effective than the region of acceptance given as a result of the PRP, that is, independently of
the probability distribution of the observable.

Suppose that $\poit{x}{m_j}, j=0,1$ are two arbitrary probability distributions indexed by
the parameters $m_0,m_1$, the latter indicating the probability distribution of term frequency
in non-relevant documents and in relevant documents, respectively.

\begin{theorem}
  For every $\poit{x}{m_j}, j=0,1$ and $x_\alpha$ 
  \begin{equation}
    \label{eq:23}
    Q_c \geq P_c \qquad Q_e \leq P_e
  \end{equation}
\end{theorem}
\begin{proof}
  Consider Figures~\ref{fig:geometry-a} and~\ref{fig:geometry-b}.  A probability of
  detection $p_d$ and a probability of false alarm $p_0$ defines the coordinates of
  $\ket{m_0}$ and $\ket{m_1}$ with a given orthonormal basis $\ket{e_0}, \ket{e_1}$ (that
  is, an observable):
  \begin{eqnarray}
    \ket{m_0} = \sqrt{1-p_d}\ket{e_0}+\sqrt{p_d}\ket{e_1}
    \\
    \ket{m_1} = \sqrt{p_0}\ket{e_0}+\sqrt{1-p_0}\ket{e_1}
    \label{eq:22}
  \end{eqnarray}
  The coordinates are expressed in terms of angles:
  \begin{equation}
    1-p_d=\sin^2\eta_1 \qquad p_0 = \sin^2\eta_0
  \end{equation}
  provided that $\eta_i$ is the angle between $\ket{m_i}$ and $\ket{e_1}$.  

  The probability of error is
  \begin{equation}
    p_e = \xi p_0 + (1-\xi)(1-p_d) = \xi \sin^2\eta_0 + (1-\xi)\sin^2\eta_1
  \end{equation}
  The probability of error is minimum when $\eta_0=\eta_1=\theta$ as shown in~\cite[page
  99]{Helstrom76}.

  But, $\theta$ is exactly the angle between $\ket{m_i}, i=0,1$ and $\ket{\mu_i}$
  and is defined as a result of Equation~\eqref{eq:1}.  The probability of error is then minimized
  when the observable vectors are the $\ket{\mu_i}, i=0,1$.

  Therefore, $Q_e \leq P_e$ for all $P_e$, that is, for all the observable vectors. As
  $Q_c = 1-Q_e$, the probability of detection is also maximum.
\end{proof}


Suppose, for example, that $X$ is a binary observable, that is, $x \in \{0,1\}$. The
probability distributions are in Table~\ref{tab:example-0}.  Suppose that the size of the
test is $\alpha=\frac{1}{5}$. Thus, relevance is accepted when $X=1$, $P_d = \frac{4}{5}$,
$P_0 = \frac{1}{5}$ and, if the states are equiprobable, $P_e = \frac{1}{2}\left(P_0 +
  1-P_d\right) = \frac{1}{5}$ and $P_c = \frac{1}{2}\left(1-P_0 + P_d\right) =
\frac{4}{5}$.

The optimal observable vectors are
\begin{equation}
  \label{eq:30}
  \ket{\mu_1} = 
  \left(
    \begin{array}{r}
      0.97 \\ -0.23
    \end{array}
  \right)
  \qquad
  \ket{\mu_0} = 
  \left(
    \begin{array}{r}
      0.23 \\ 0.97
    \end{array}
  \right)
\end{equation}
These vectors can be computed in compliance with~\cite{Eldar&01}.  Hence, the region of
acceptance is the subspace spanned by $\ket{\mu_1}$ and, if the states are equiprobable,
$Q_e = \frac{1}{2}\left(Q_0 + 1-Q_d\right) = 0.05$ and $Q_c = \frac{1}{2}\left(1-Q_0 +
  Q_d\right) = 0.95$. Hence, if we were able to find the optimal observable vector and to
actually measure it, retrieval performance would be much higher than the performance
achieved through the classical region of acceptance.

\begin{figure}[t]
  \centering
  \begin{algorithmic}
    \STATE sort data by increasing $f_w(x;m)$ and by $m$
    \FORALL{topic $t$}{
      \STATE extract title and description of $t$
      \FORALL{topic word $w$ of $t$}{
        \STATE compute $\poitw{w}{x}{m}$ for every $x$
        \STATE compute $\interm$
        \FORALL{size $\alpha \in \left\{0.25, 0.50, 0.75\right\}$}{
          \FORALL{prior $\xi \in \left\{0.01, \ldots, 0.99\right\}$}{
            \STATE compute $P_e, P_c$
            \STATE compute $Q_e, Q_c$
          }
          \ENDFOR
        }
        \ENDFOR
      }
      \ENDFOR
    }
    \ENDFOR
  \end{algorithmic}
  \caption{The experimental algorithm}
  \label{fig:experiment-0}
\end{figure}




\section{Experimental Study}
\label{sec:experiments}

We have tested the theory illustrated in the previous sections through experiments based
on the TIPSTER test collection, disks 4 and 5. The experiments aimed at measuring the
difference between $P_e$ and $Q_e$ by means of a realistic test collection. To this end, we
have used the TREC-6, 7, 8 topic sets.  The experimental algorithm is explained in
Figure~\ref{fig:experiment-0}. 

We have implemented the following test: $\poit{x}{m}$ has been computed for each topic
word and for each $m$ by means of the usual relative frequency $f_w(x;m)/\sum_x f_w(x;m)$
assuming that $f_w(x;m)$ is the frequency of $w$ in the documents with state $m$. Note
that we do not aim at measuring effectiveness; rather, we aim at measuring the difference
between probabilities of error \emph{given} a document ranking.

\begin{figure}[ht]
  \centering
 \subfigure[$\alpha=0.25$]{
    \epsfig{width=0.3\columnwidth,file=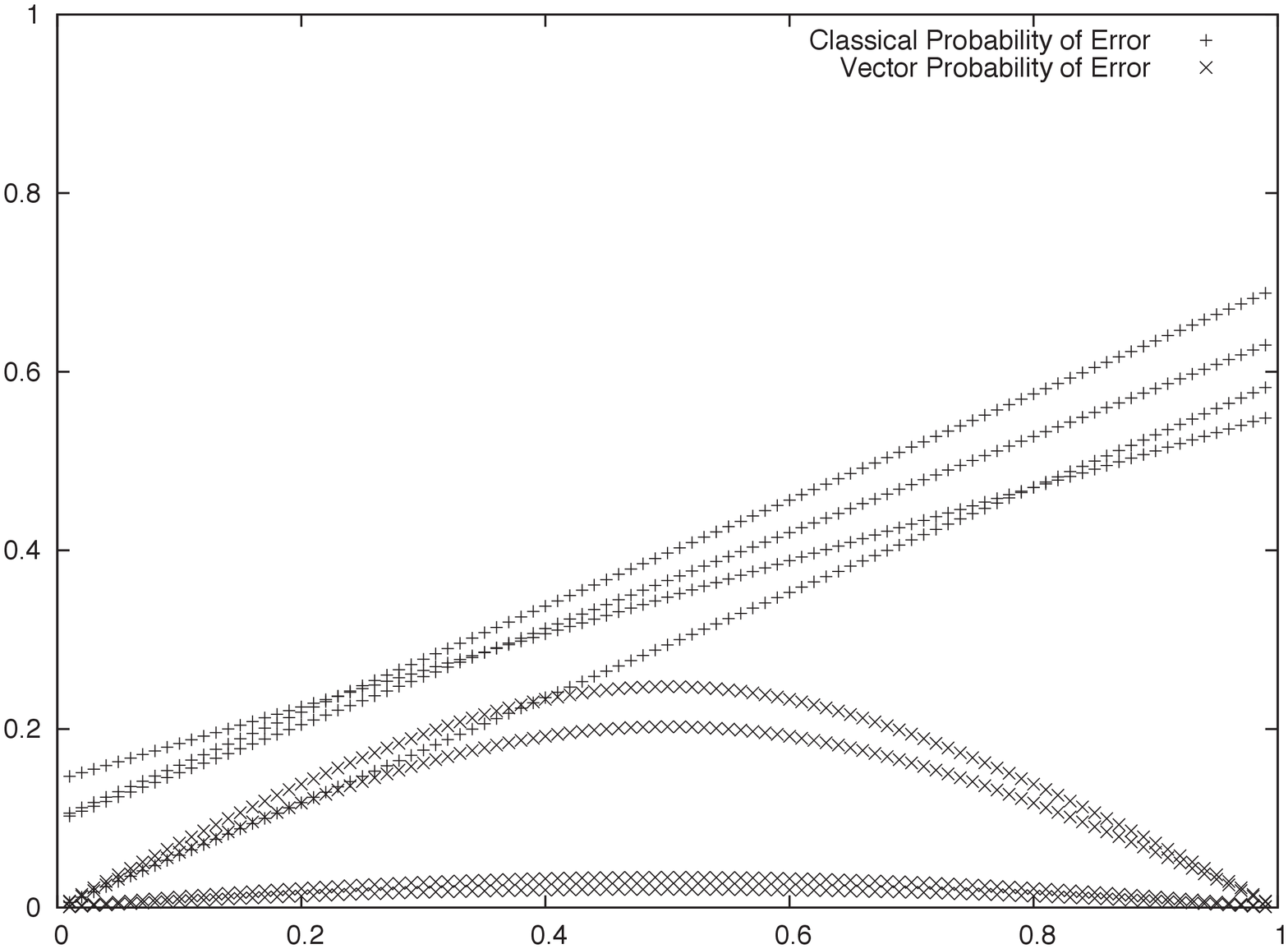}
  }
  \subfigure[$\alpha=0.50$]{
    \epsfig{width=0.3\columnwidth,file=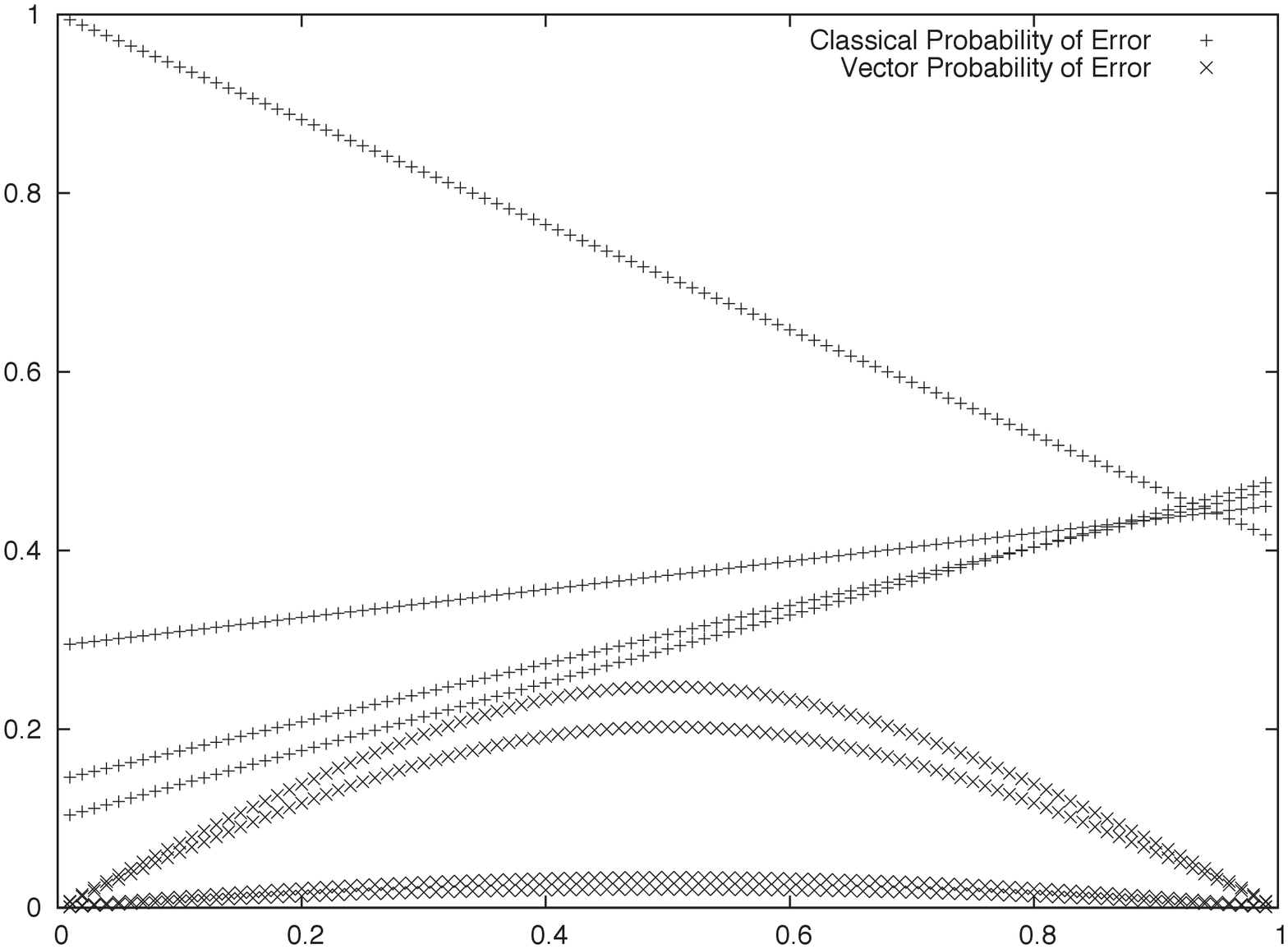}
  }
  \subfigure[$\alpha=0.75$]{
    \epsfig{width=0.3\columnwidth,file=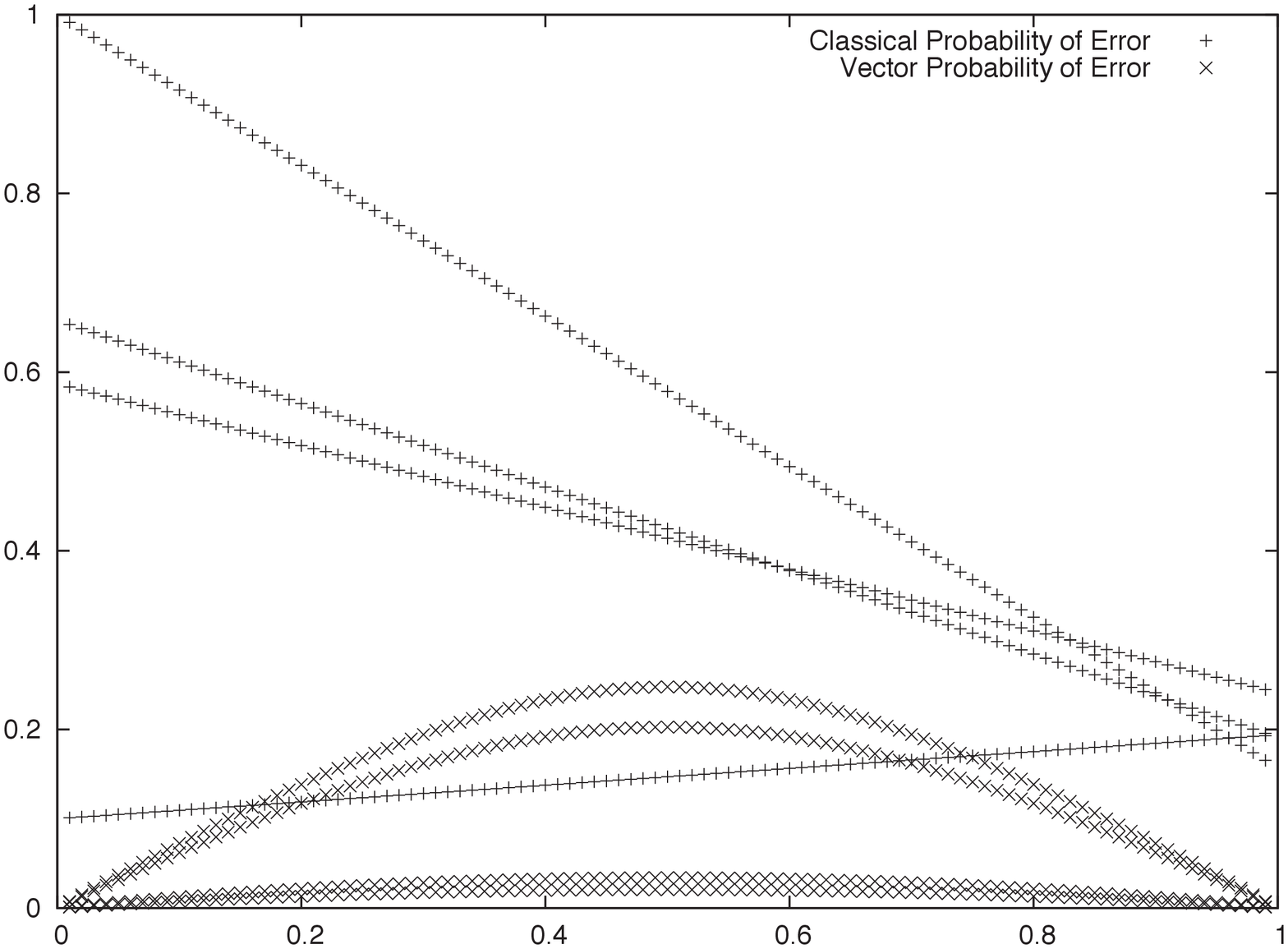}
  }
  \caption{$P_e$ and $Q_e$ plotted against $\xi$ for each word of topic $439$ and for each
    $\alpha \in \left\{0.25, 0.50, 0.75\right\}$; each curve corresponds to a word:
    \textsf{+} labels classical probability of error curves, $\mathsf{\times}$ labels vector
    probability of error curves.}
  \label{fig:output-t439}
\end{figure}

Consider Figure~\ref{fig:output-t439}: $Q_e$ is always not greater than $P_e$ for every
size $\alpha$ and for every prior probability $\xi$. The superposition of linear curves,
one curve for each $\alpha$, yields a polygonal curve like Figure~\ref{fig:polygonal}.

Some linear curves are secant because they cut a bell-shaped curve in two parts. However,
they refer to different words: given a word, the linear curve is never secant of the
bell-shaped curve.

\begin{figure*}[ht]
  \centering
  \subfigure[$\alpha=0.25$]{
    \epsfig{width=0.3\columnwidth,file=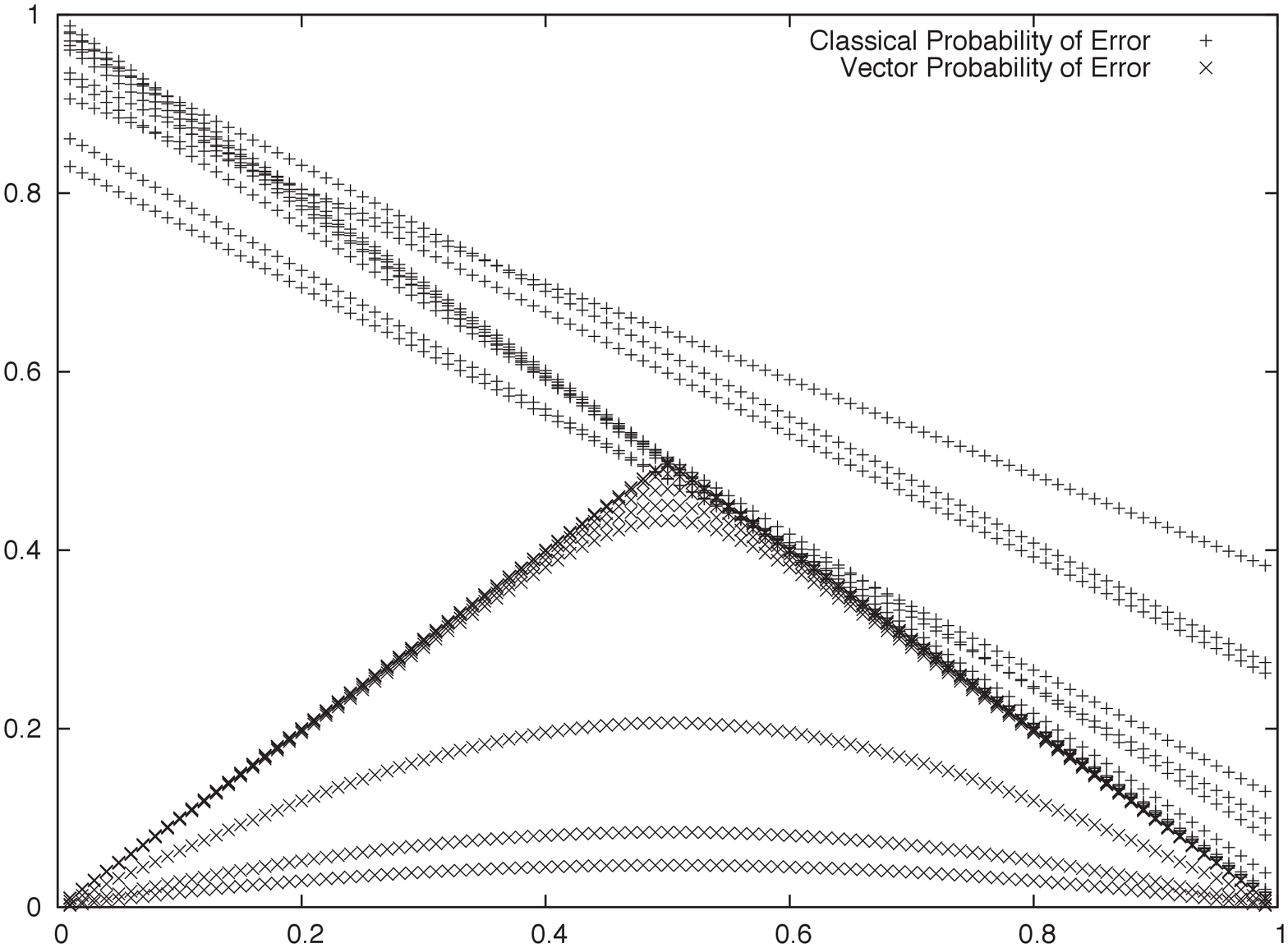}
      \label{fig:output-t301-a025}
  }
  \subfigure[$\alpha=0.50$]{
    \epsfig{width=0.3\columnwidth,file=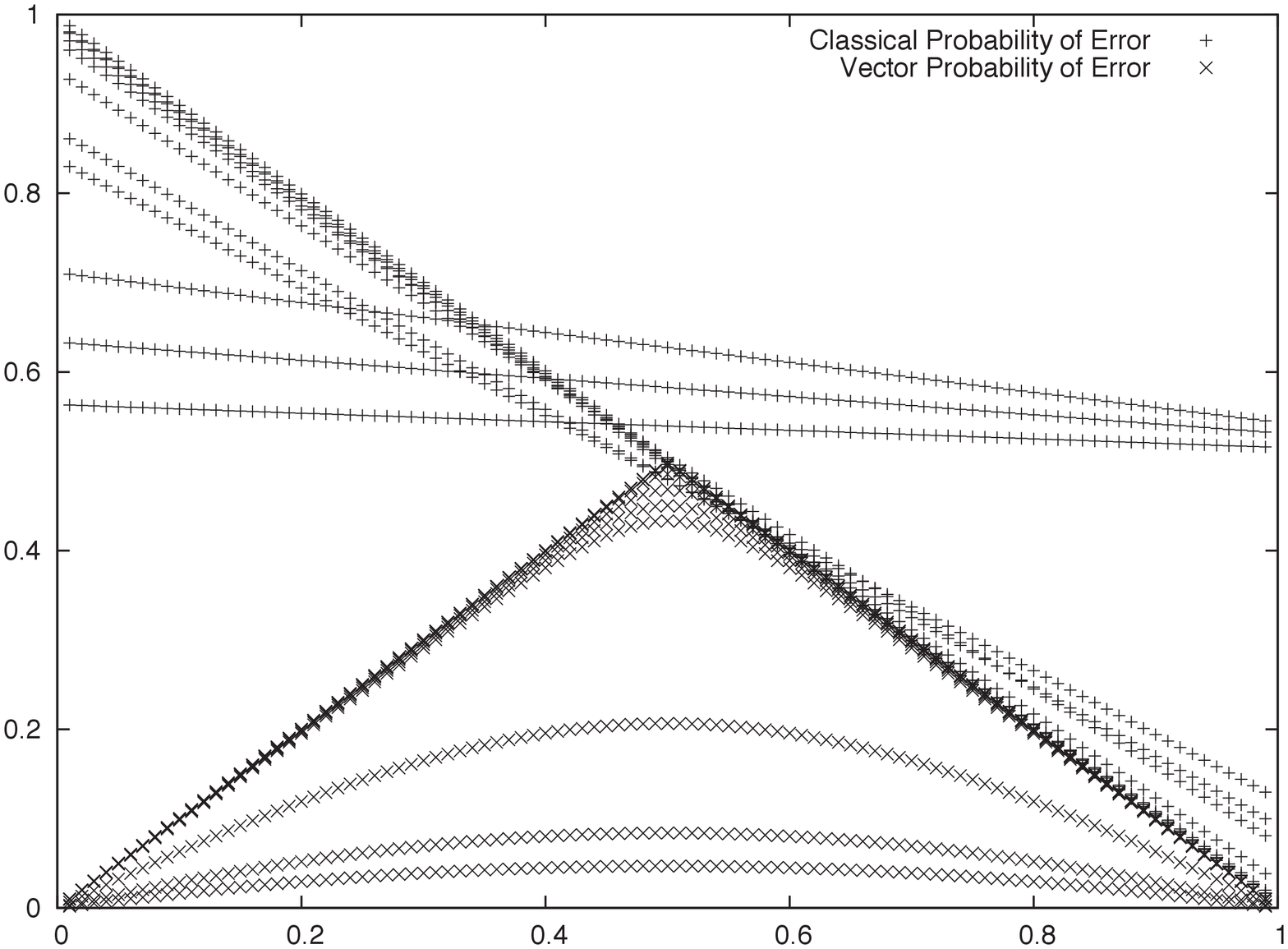}
      \label{fig:output-t301-a050}
  }
  \subfigure[$\alpha=0.75$]{
    \epsfig{width=0.3\columnwidth,file=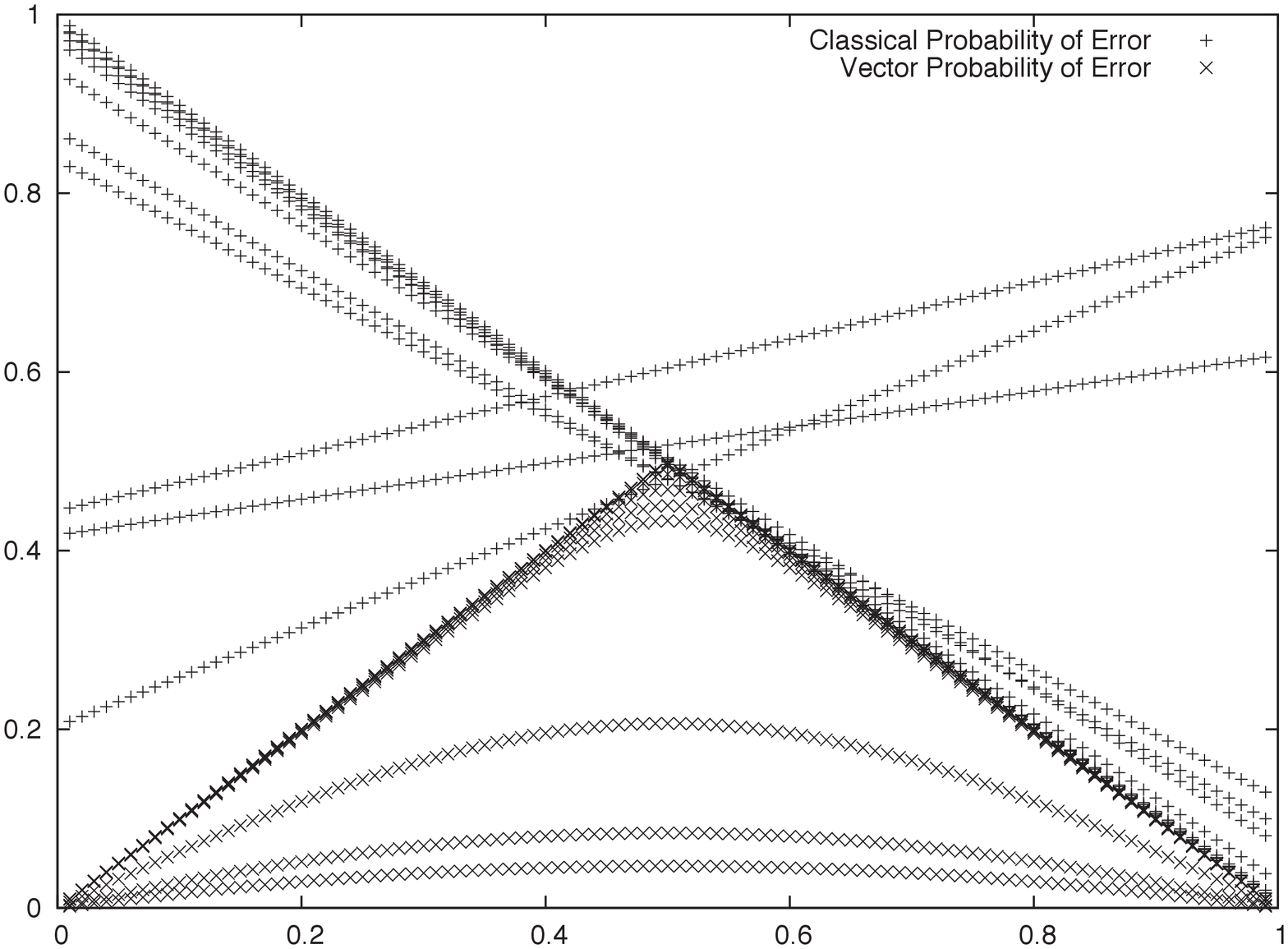}
      \label{fig:output-t301-a075}
  }
  \caption{$P_e$ and $Q_e$ plotted against $\xi$ for each word of topic $301$
    (Figures~\ref{fig:output-t301-a025},~\ref{fig:output-t301-a050}
    and~\ref{fig:output-t301-a075})) and for each $\alpha \in \left\{0.25, 0.50,
      0.75\right\}$; each curve corresponds to a word: \textsf{+} labels classical
    probability of error curves, $\mathsf{\times}$ labels vector probability of error
    curves.}
  \label{fig:output-t301}
\end{figure*}

\begin{figure*}[ht]
  \centering
  \subfigure[$\alpha=0.25$]{
    \epsfig{width=0.3\columnwidth,file=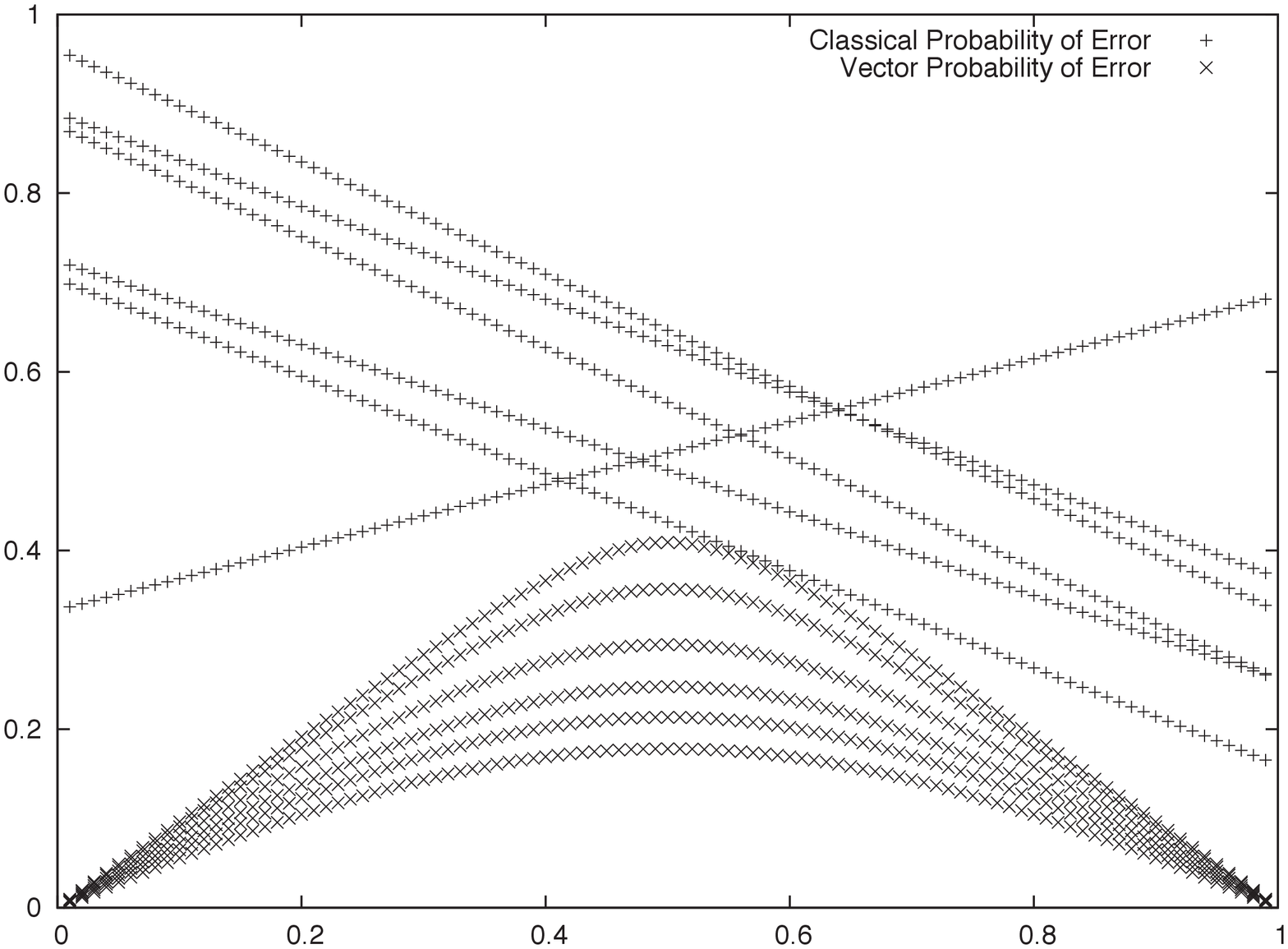}
      \label{fig:output-t303-a025}
  }
  \subfigure[$\alpha=0.50$]{
    \epsfig{width=0.3\columnwidth,file=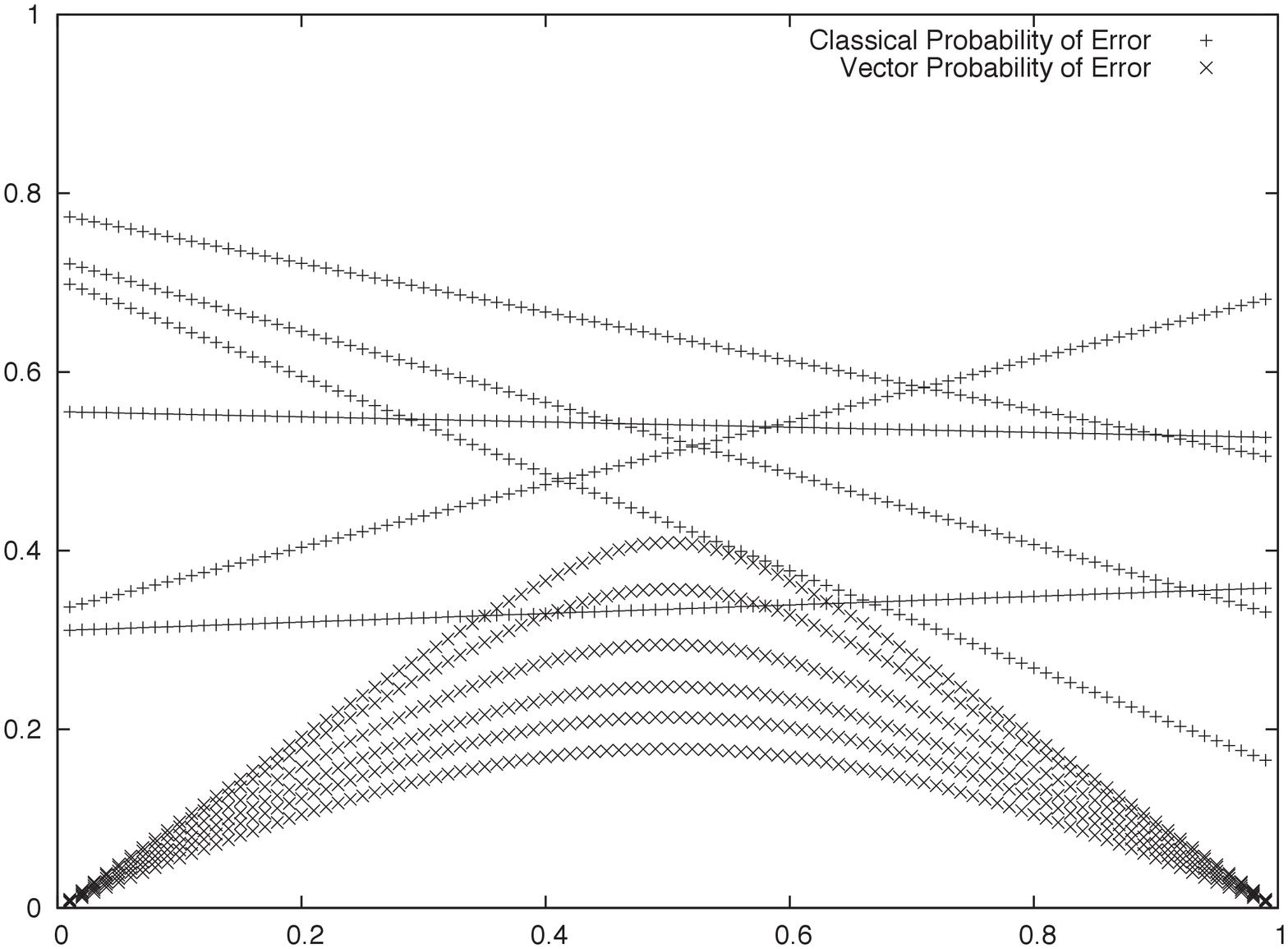}
      \label{fig:output-t303-a050}
  }
  \subfigure[$\alpha=0.75$]{
    \epsfig{width=0.3\columnwidth,file=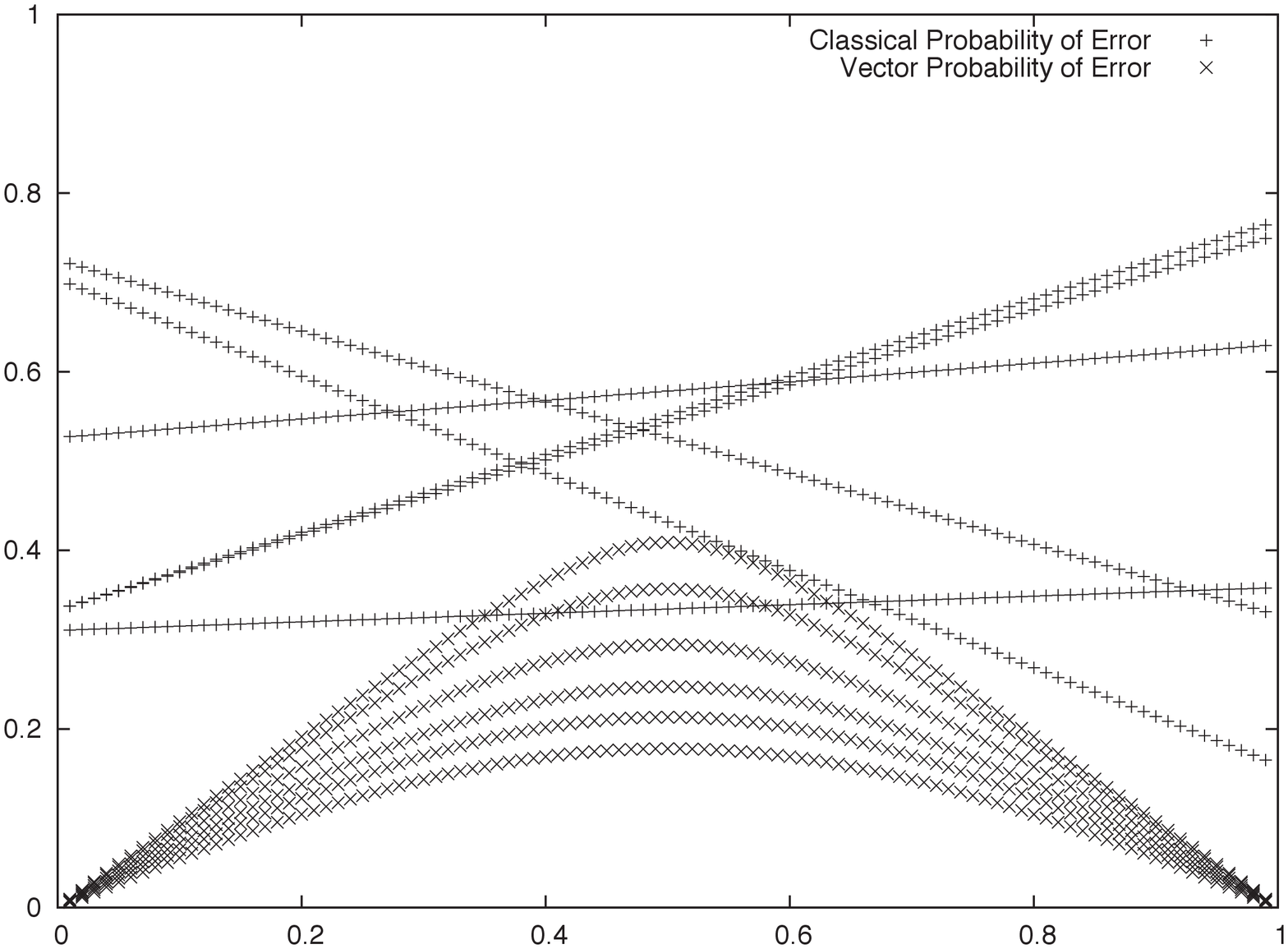}
      \label{fig:output-t303-a075}
  }
  \caption{$P_e$ and $Q_e$ plotted against $\xi$ for each word of topic $303$
    (Figures~\ref{fig:output-t303-a025},~\ref{fig:output-t303-a050}
    and~\ref{fig:output-t303-a075})) and for each $\alpha \in \left\{0.25, 0.50,
      0.75\right\}$; each curve corresponds to a word: \textsf{+} labels classical
    probability of error curves, $\mathsf{\times}$ labels vector probability of error
    curves.}
  \label{fig:output-t303}
\end{figure*}

\begin{figure*}[ht]
  \centering
  \subfigure[$\alpha=0.25$]{
    \epsfig{width=0.3\columnwidth,file=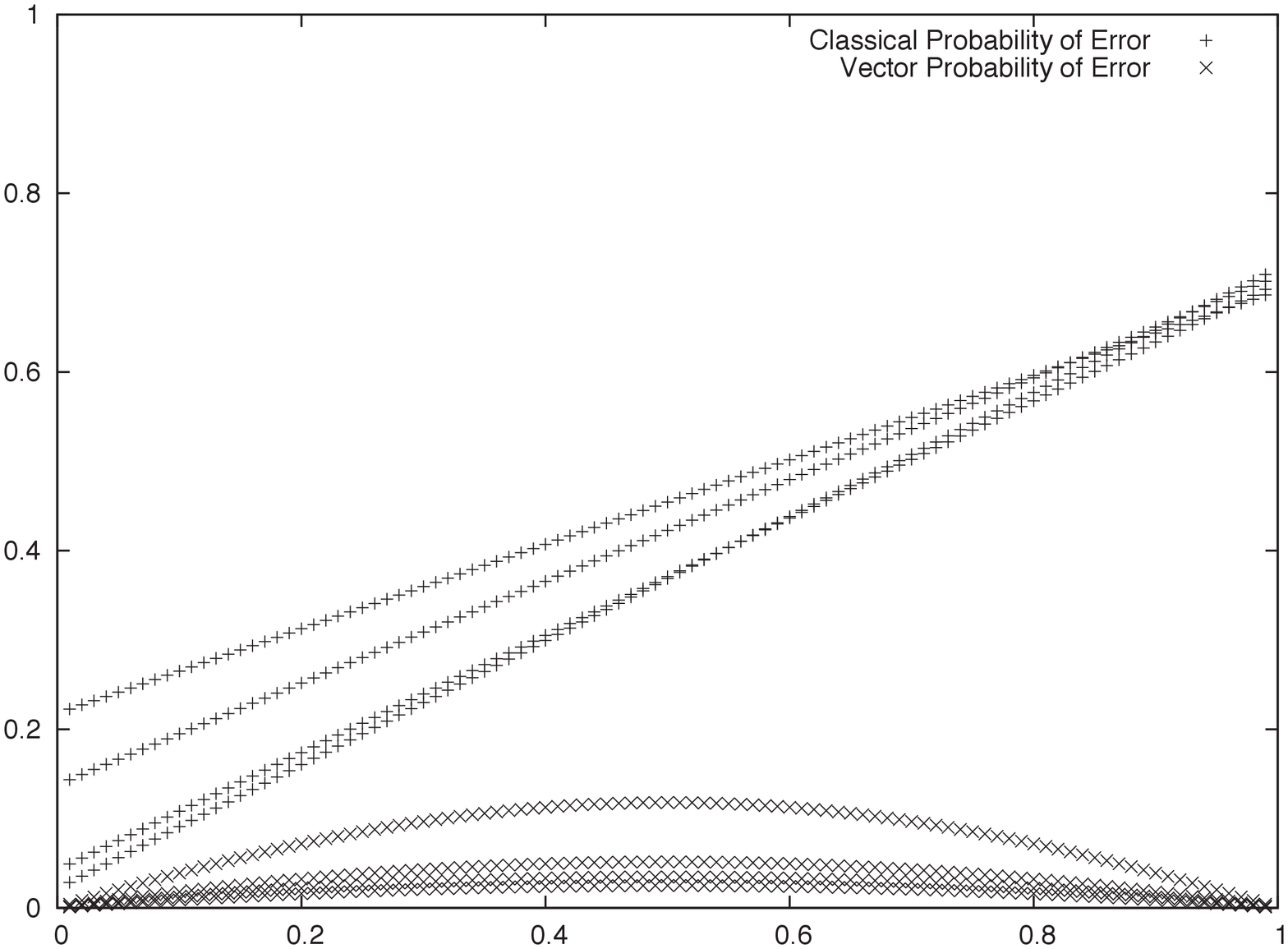}
      \label{fig:output-t392-a025}
  }
  \subfigure[$\alpha=0.50$]{
    \epsfig{width=0.3\columnwidth,file=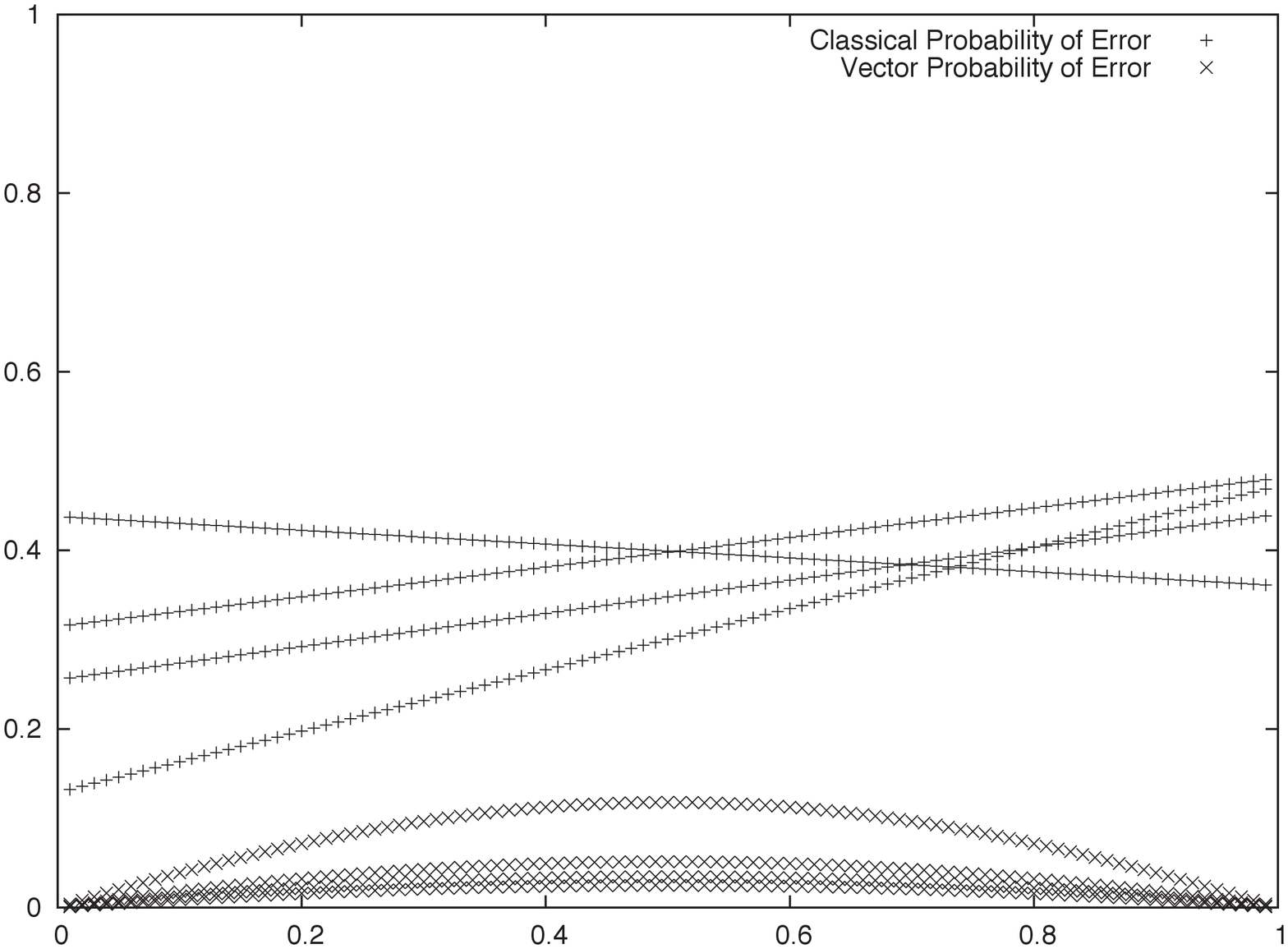}
      \label{fig:output-t392-a050}
  }
  \subfigure[$\alpha=0.75$]{
    \epsfig{width=0.3\columnwidth,file=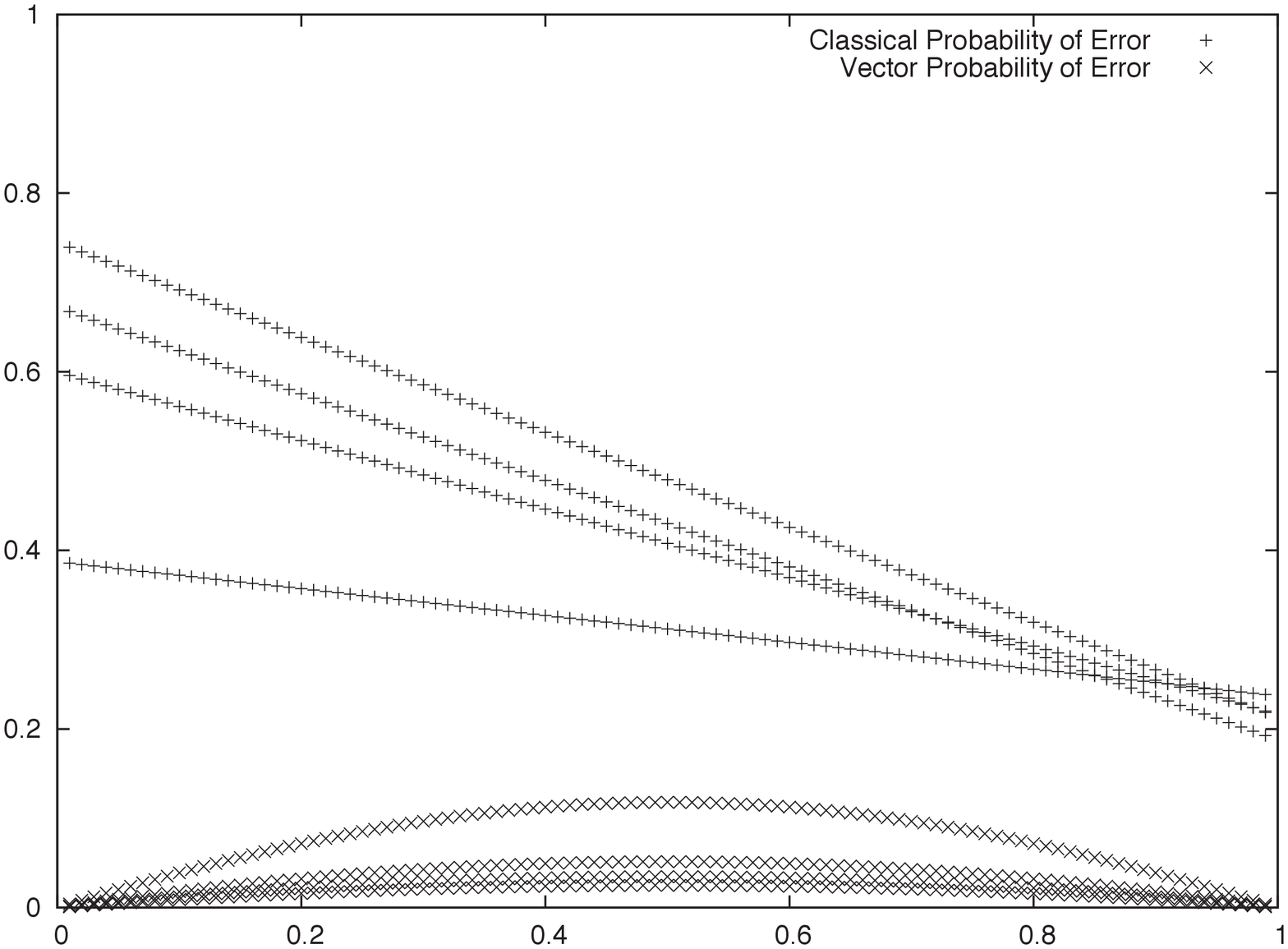}
      \label{fig:output-t392-a075}
  }
  \caption{$P_e$ and $Q_e$ plotted against $\xi$ for each word of topic $426$
    (Figures~\ref{fig:output-t392-a025},~\ref{fig:output-t392-a050}
    and~\ref{fig:output-t392-a075}) and for each $\alpha \in \left\{0.25, 0.50,
      0.75\right\}$; each curve corresponds to a word: \textsf{+} labels classical
    probability of error curves, $\mathsf{\times}$ labels vector probability of error
    curves.}
  \label{fig:output-t392}
\end{figure*}
 
Figures~\ref{fig:output-t301},~\ref{fig:output-t303},~\ref{fig:output-t392} illustrate the
plots for other three topics; these topics are representative of the main types of
plot~--~the plots of all the $450$ topics exhibit the similar pattern.

\section{Discussion}
\label{sec:discussion}

The precedent example points out the issue of the measurement of the optimal observable
vectors.  Measurement means the actual finding of the presence / absence of the optimal
observable vectors via an instrument or device.  The measurement of term frequency is
straightforward because term occurrence is a physical property measured through an
instrument or device. (A program that reads texts and writes frequencies is sufficient.)
The measurement of the optimal observable vectors is much more difficult because no
physical property does correspond to them and cannot be expressed in terms of term
frequencies.~\cite{Griffiths02} Thus, the question is: what should we observe from a
document so that the outcome corresponds to the optimal observable vector?  The question
is not futile because the answer(s) would effect automatic indexing and retrieval.  In
particular, if we were able to give an interpretation to the optimal observable vectors,
retrieval and indexing algorithms could measure those vectors.

Following~\cite{vanRijsbergen04}, three interpretations of the optimal observable vectors
can be provided:
\begin{figure}[t]
  \centering
  \setlength{\unitlength}{1mm}
  \begin{minipage}[t]{1.0\columnwidth}
    \begin{picture}(50,60)(-20,-25)
      \put(0,0){\vector(1,0){35}}\put(1,33){$\ket{1}$}
      \put(0,0){\vector(0,1){35}}\put(33,1){$\ket{0}$}
      \put(0,0){\vector(1,+1){25}}\put(25,22){$\ket{\mu_1}$}
      \put(0,0){\vector(1,-1){25}}\put(25,-23){$\ket{\mu_0}$}
    \end{picture}
  \end{minipage}
  \caption{A geometric view of incompatible observable vectors}
  \label{fig:geometry-c}
\end{figure}
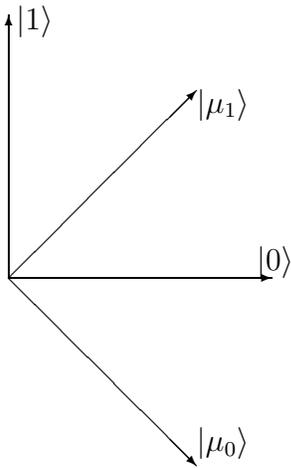

\begin{itemize}
\item Geometrically, each vector is a superposition of other two independent vectors.
  Figure~\ref{fig:geometry-c} depicts the way the vectors interact and shows that the
  observation of a binary feature places the observer upon either $\ket{0}$ or $\ket{1}$
  whenever he measure $0$ or $1$, respectively. There is no way to move upon $\ket{\mu_0}$
  or $\ket{\mu_1}$ because $\mu_0, \mu_1$ cannot be measured.  
\item Probabilistically, the observable vectors and the state vectors are related as a result of the
  rule of Equation~\ref{eq:9}. (Also known as trace rule because the general form of the
  equation is the trace of two matrices.) As the observable vectors are mutually
  orthonormal by definition, they induce a valid probability distribution.  
\item Logically, the observable vectors are assertions, e.g., $X=x$ corresponds to
  $\ket{x}$. The basic difference between subspaces and subsets is that the vectors belong
  to a subspace if and only if they are spanned by a basis of the subspace. However, the
  logic to combine subspaces cannot be the set-based logic used to combine subsets.
\end{itemize}
In classical probability theory, if we observe $1$, we say that every document described
by $1$ is either relevant or not relevant, when relevance is measured. In general, we say
that it either possesses a property or does not, when a property is measured. Hence, if an
observable is described as sets of values (e.g., the set of documents indexed by a term),
we can always describe relevance as a set. That is, the union of the set of relevant
documents indexed by a term with the set of relevant documents not indexed by the term.

The orthogonality between the observable vectors implements the mutually exclusiveness
between the observable values. Hence, if we observe $0$, we can say only that we do not
observe $1$, but cannot say anything about $\mu_1$ because $\ket{\mu_1}$ is oblique to
$\ket{0}, \ket{1}$ and vector subspace complement, union and intersection are not the same
as subset complement, union and intersection~\cite{vanRijsbergen04}.

At the present time, an IR system is capable of measuring observable vectors like
$\ket{0}, \ket{1}$ which correspond to term occurrence.  The documents can be ranked as
specified by the PRP, thus achieving $P_e$, which is the current lower bound provided that
the probabilities are estimated as accurately as possible~\cite{Robertson77}.

$Q_e$ and $Q_c$ can be achieved \emph{if and only if} an IR system is capable of observing
the optimal observable vectors (Theorem~\ref{the:helstrom}).  If an IR system observed
$\mu_0$ or $\mu_1$ in a document, the system would decide whether the document is relevant
with probability of error $Q_e$.

The open problem is due to the difficulty of observing the optimal observable vectors in a
document; if a system is given a textual document as input, how can it decide if the
document would provide either $\ket{\mu_0}$ or $\ket{\mu_1}$ (or the corresponding
eigenvalues) if the optimal observable vectors were measured? We shall pay a great deal of
attention to the question because, if the problem were solved, the solution would give a
significant breakthrough in IR research.

\section{Related Work}
\label{sec:related-work}

Van Rijsbergen's book~\cite{vanRijsbergen04} is the point of departure of our work.
Helstrom's book~\cite{Helstrom76} provides the theoretical foundation for the vector
probability and the optimal observabl vectors.  Eldar and Forney's paper~\cite{Eldar&01}
gives an algorithm for the optimal observable vectors.  Hughes' book~\cite{Hughes89} is an
excellent introduction to Quantum Theory.  An introduction to Quantum Theory and
Information Retrieval can be found in~\cite{Melucci&11}.  In~\cite{Piwowarski&10b} the
authors propose quantum formalism for modeling some IR tasks and information need aspects.
The paper does not limit the research to the application of an abstract formalism, but
exploits the formalism to illustrate how the optimal observable vectors significantly
improve effectiveness.  In~\cite{Zuccon&10}, the authors propose $\interm$ for modifying
probability of relevance; $\interm$ is intended to model quantum correlation (also known
as interference) between relevance assessments.

\section{Conclusions}
\label{sec:conclusions}

The research in IR has been traditionally concentrated on extracting and combining
evidence as accurately as possible in the belief that the observed features (e.g., term
occurrence, word frequency) have to ultimately be scalars or structured objects.  The
quest for reliable, effective, efficient retrieval algorithms requires to implement the
set of features as best one can.  The implementation of a set of features is thus an
``answer'' to an implicit ``question'', that is, which is the best \emph{set} of features
for achieving effectiveness as high as possible?  We suggest to ask another ``question''
to achieve an even better answer: Which is the best \emph{subspace}?

\end{document}